\documentclass[final,5p,times,twocolumn]{elsarticle}

\usepackage{graphicx}
\usepackage{amssymb}
\usepackage{amsthm}

\usepackage{lineno}

\usepackage{amsmath}

\usepackage{multirow}
\usepackage{subfig}
\usepackage{float}
\usepackage{amsmath}
\newcommand{\abs}[1]{\left\lvert #1 \right\rvert}

\usepackage{romannum}
\usepackage{color,soul}
\usepackage{algorithm}
\usepackage{algorithmic}

\journal{Computers \& Chemical Engineering}

\begin{document}

\begin{frontmatter}


\title{Twin actor twin delayed deep deterministic  policy  gradient (TATD3) learning for batch process control }


\author{Tanuja Joshi$^a$}
\author{Shikhar Makker$^a$}
\author{Hariprasad Kodamana$^{a,b}$}
\ead{kodamana@iitd.ac.in}
\author{Harikumar Kandath$^c$}
\ead{harikumar.k@iiit.ac.in}
\address{$^a$Department of Chemical Engineering,
Indian Institute of Technology Delhi,
Hauz Khas, New Delhi - 110016\\
$^b$School of Artificial Intelligence,
Indian Institute of Technology Delhi,
Hauz Khas, New Delhi - 110016\\
$^c$International Institute of Information Technology Hyderabad,
Gachibowli, Hyderabad - 500 032, India}

\begin{abstract}

Control of batch processes is a difficult task due to their complex nonlinear dynamics and unsteady-state operating conditions within  batch and batch-to-batch. It is expected that some of these challenges can be addressed by developing control strategies that directly interact with the process and learning from  experiences. Recent studies in the literature have indicated the advantage of having an ensemble of actors in actor-critic Reinforcement Learning (RL) frameworks for improving the policy. The present study proposes an actor-critic RL algorithm, namely, twin actor twin delayed deep deterministic policy gradient (TATD3), by incorporating twin actor networks in the existing twin-delayed deep deterministic policy gradient (TD3) algorithm for the continuous control.  In addition, two types of novel reward functions are also proposed for TATD3 controller. We showcase the efficacy of the TATD3 based controller for various batch process examples  by comparing it with some of the existing RL algorithms presented in the literature.

\end{abstract}


\end{frontmatter}


\section{Introduction}
\label{S:1}
 Studies have emphasized the usefulness of batch processes in industrial-scale production of various value-added chemicals because of advantages like low capital cost, raw material cost, and flexibility in operation \cite{hosseini2012biodiesel}. However, the operation of the batch process grapples with several challenges due to the characteristics such as non-linearity due to the temperature-dependent kinetics, time-varying dynamics and a broad range of operating conditions. Therefore, the control and optimization of the batch process is a very challenging task. In most of the batch processes, reaction temperature has a significant effect on the yield and, therefore, the primary control variable in the existing literature on the control of the batch process. \cite{chanpirak2017improvement,benavides2012optimal,De2016DynamicProduction}. 
In the end, the yield of the product depends mostly on the choice of the controller and its capability and the tuning parameters. Even if the controller is appropriately tuned for an operating condition, the drifts in process parameters can significantly deteriorate the controller performance over time.
To overcome some of these issues, advanced control strategies such as model predictive control (MPC), iterative learning control (ILC), non-linear MPC (NMPC) are presented in the literature for the  control of the batch process \cite{marchetti2016modifier,mesbah2016stochastic,lee2007iterative,Kern2015AdvancedReactor,mjalli2009approximate,mjalli2009dynamics,kuen2010recursive,benavides2012optimal,brasio2013nonlinear,brasio2016first}.

The performance of a model-based controller is heavily dependent on the accuracy of the underlying model of the process. Even the slightest inaccuracies in the model will result in plant-model mismatch and, consequently, inaccurate prediction of the concentration profiles of the species from the model. These issues could arise mainly due to batch-to-batch variations or due to parameter drifts. To overcome some of the limitations of the first-principle based model, data-driven modelling approaches for nonlinear dynamic batch processes are reported in the literature \cite{joshi2020novel,liu2008transesterification,jiang2019data}.
Further, optimization-based control algorithms are computationally demanding as the computation of the optimal input sequence involve online optimization at each time step. Despite the advances in numerical methods and computational hardware, this is still a challenging task for complex nonlinear, high-dimensional dynamical systems.
Hence, there is much incentive if a control strategy can directly interact with process trajectory and provide a control solution for online course correction by learning the operational data profile.

To this extent, it is useful to explore the feasibility of reinforcement learning (RL) as a  potential paradigm for controlling batch processes.  Contrary to classical controllers, RL based controllers do not require a process model or control law; instead, they learn the dynamics of the process by directly interacting with the operational environment \cite{spielberg2020deep,nian2020review,kim2020model}. As a result, the controller performance will not be contingent on having a  pressing requirement of a high fidelity model of the process. Further, as opposed to traditional controllers, the RL-based controller learns from experiences and past history, thus improving the control policy at every step. As an earlier proposition in this direction,  approximate dynamic programming (ADP) approach that learns the optimal `cost-to-go' function has been proposed for the optimal control of non-linear systems \cite{lee2005approximate,Lee2004AControl,Lee2010ApproximateControl}. Following this, an ADP based `Q-learning' approach that learns the optimal policy using value iteration in a model-free manner has also been proposed  \cite{lee2005approximate}.
Unlike the traditional applications of `Q-learning',  the state-space and action-space are continuous for most process control applications. This poses a major limitation to the application of conventional Q-learning methods in this context. For systems with continuous state-space, Mhin et al. combined Q-learning with a deep neural network for function approximation and developed a deep Q network (DQN) \cite{mnih2015human}. 
Inspired by these, some of the recent works related to RL applications in chemical processes have attempted to optimize the control policy by applying  Q-learning and deep Q-learning for applications such as chromatography and polymerisation. \cite{nikita2021reinforcement, singh2020reinforcement}.

Although the advent of DQN based RL was a major breakthrough in applications to systems with continuous state-space, their utility to systems with continuous action-space was still limited.   Another approach to solving the RL problem is by employing policy gradient (PG), where instead of evaluating the value functions to find an optimal policy, the optimal policy is evaluated directly.  This approach is particularly well-suited to deal with problems where both the state and action spaces are continuous.  For instance, Petsagkourakis et al. applied the PG algorithm to find the optimal policy for a batch bioprocess by using principles of transfer learning \cite{Petsagkourakis2020ReinforcementOptimization}. However, PG methods suffer from the drawback of noisy gradients due to high variance, leading to slow convergence.  Actor-critic methods reduce this low variance gradient estimates by exploiting a critic network and has been the widely used framework for dealing with continuous action spaces. In this connection, Deep Deterministic Policy Gradient (DDPG)is one of the actor-critic algorithms that has celebrated a huge success \cite{Lillicrap2015Continuous}. Indeed,  DDPG was the first efficient algorithm used to solve high dimensional continuous control tasks.  It effectively combines the architecture of DQN and deterministic policy gradient (DPG).{ Recent works} have shown the application of the DDPG algorithm for chemical process control as well. For example, Ma et al. have proposed a DDPG based controller for the control of semi-batch polymerisation \cite{Ma2019ContinuousLearning}, Spielberg et al. has applied the DDPG algorithm for SISO and MIMO linear systems \cite{spielberg2020deep}. Recent work by Yoo et al. proposed a modified DDPG based controller for stable learning and reward function design for the control of batch polymerisation process \cite{yoo2021reinforcement}. For a detailed review of the application of RL for chemical control problems, the readers are referred to \cite{spielberg2019toward,nian2020review,shin2019reinforcement}.

Fujimoto et al. \cite{fujimoto2018addressing} have shown that the DDPG algorithm suffers from a critical issue of overestimation of network bias due to function approximation error and leads to sub-optimal policy. The authors have also provided an approach to address the function approximation error in actor-critic methods and termed the new algorithm as Twin Delayed Deep Deterministic Policy Gradient (TD3). Some recent work involving the application of TD3 algorithm involves motion planning of robot manipulators, half cheetah robot as an intelligent agent to run across a field, etc. \cite{kim2020motion,dankwa2019modeling}. Further, some of the recent studies in the literature have also indicated the advantage of having an ensemble multiple actors in actor-critic RL frameworks for improving optimal policy. In a multi-actor architecture, overall policy is obtained by amalgamating the results of parallel training of multi-actor networks \cite{liu2020reinforcement}.

Inspiring from these, this paper presents a novel actor-critic framework termed as twin actor twin delayed deep deterministic  (TATD3) policy gradient learning, combining TD3 and multi-actor frameworks, for the control of the batch process. 

We also propose novel reward functions analogous to proportional-integral (PI) and proportional-integral-derivative (PID) functions, and their performances are compared. We have also compared the results of the TATD3 approach with popular actor-critic methods and RL algorithms with discrete action spaces. The performance of the different algorithms is compared by numerical evaluation of two batch processes viz.  (i) batch-transesterification process, and (ii) exothermic batch process. TATD3 algorithm-based controller shows better performance in terms of tracking error and control effort when compared to TD3 and DDPG algorithm based controllers for both of the batch processes.
We believe that the following are the novel components of this study.
To the best of our knowledge (i) amalgamation of twin actors in a TD3 framework  has not been reported in literature for process control;  (ii) we showcase that reward functions inspired based on PI and PID controller functions works well in an RL framework, (iii) we also validate the efficacy of the proposed TATD3 algorithm by comparing the results with TD3, DDPG and other RL algorithms based on discrete action space. Such a comprehensive study involving the comparison of  RL based algorithms with continuous and discrete action spaces has also not been reported in the literature in the context of the process control.

The rest of the paper is divided into the following sections. Section 2 presents the background of RL useful for developing a TATD3 based controller.
Section 3 explain the TATD3 algorithm in detail.
Section 4 shows the application of the proposed controller and important results  for the control of (i) batch-transesterification process and (ii) exothermic batch process. Section 5 draws conclusive remarks from the study.

\section{Background}
\subsection{Q- learning and deep Q-learning}

A standard RL framework consists of three main components, namely the agent, the environment ($E$) and the reward ($r$) \cite{sutton2018reinforcement}. 
At each time step $t$, the agent performs an action $a_{t}$ based on the  state $s_{t}$ of the environment and receives a scalar reward $r_{t}$, 
as a result,  the environment moves to a new state $s_{t+1}$.
The objective of the RL problem is to find the sequence of control actions $\mathcal{A}:=\{a_0,\,a_1,\,a_2,\dots\}$ to maximise the expected discounted reward as given below:
\begin{equation}
    \arg \underset{\mathcal{A}}{ \max}\,\,\, \mathbb{E}~[R_{t} :=\sum\limits_{k=0}^\infty   \gamma^{k}{r_{t+k}}]\label{eq:rl}
\end{equation}

where $0<\,\gamma\,<1$ is the discount factor and $\mathbb{E}$ denotes the expectation operator which is applied to the discounted reward due to the
 stochastic nature of the process dynamics.
However, the explicit solution of Eq.\eqref{eq:rl} is tedious to obtain.  Q-learning is an iterative algorithm to solve the RL problem over a finite set of actions.
The Q-value of a state-action pair, $Q_{\pi}(s_t,a_t)$ is the expected return after performing an action $a_t$ at a state $s_t$  following a policy $\pi: \mathcal{S} \rightarrow \mathcal{A}$ :
\begin{equation}
    Q_{\pi}(s_t,a_t)= \mathbb{E}_{\pi}\Big[\sum\limits_{k=0}^\infty   \gamma^{k}{r_{t+k}}|s_{t}=s,a_{t}=a\Big] 
\end{equation}
where $\mathcal{S}=\{s_0,s_1,s_2,\dots\}$.
The objective of the Q-learning is to find the optimal policy ($\pi^{*}$) by learning the optimal Q-value, $Q^{*}(s_t,a_t)$ which is the maximum expected return achievable by any policy for a state-action pair.

The optimal Q-value, $Q^{*}(s_t,a_t)$ must satisfy the Bellman optimality equation \cite{barron1989bellman} given as :
\begin{equation}
   Q^{*}(s_t,a_t)= \mathbb{E}[r_{t}+\gamma \max_{a_{t+1}}Q^{*}(s_{t+1},a_{t+1}|s_{t}=s,a_{t}=a)]
\end{equation}
where $s_{t+1}$ and $a_{t+1}$ are the state and action at the next time step.
The Q-learning algorithm iteratively updates the Q-value for each state-action pair until the Q-function converges to the optimal Q-function. This is known as value iteration and is given as:
\begin{equation}
    Q(s_t,a_t) \leftarrow Q(s_t,a_t)+\alpha (r_t+\gamma \max_{a_{t+1}}Q(s_{t+1},a_{t+1})-Q(s_{t},a_{t}))
\end{equation}
{where $\alpha$  is the learning rate}.
\par
Major encumbrances of traditional Q-learning are:  (i) its application is limited only to the problems with discrete state and actions spaces;   
(ii)  computational difficulty faced while dealing with large state space owing to the large size of the Q-matrix. The former problem can be circumvented by employing a function approximator for modelling the relation between Q-value and state-action pairs.  
Deep Q-learning \cite{mnih2015human} is an RL framework wherein a Deep neural network (DNN), termed as the value network,  is used as a function for approximating the optimal Q-values. 
To address the problem of correlated sequences, DQN uses a replay buffer or experience replay memory which has a pre-defined capacity where all the past experiences are stored as the following transition tuple ($s:=s_t,a:=a_t,s':=s_{t+1},r:=r_t)$). The DQN uses past experiences to train the policy network by selecting suitable mini-batches from the replay buffer. The state is given as input to the value network, and the network outputs the Q-value corresponding to all possible actions in the action space. The loss is calculated as the mean square error (MSE) between the current Q-value and the target Q-value as given in Eq. \eqref{EQ1}:
\begin{align}
\label{EQ1}
    Loss &= \mathbb{E}\big[\big(Q^{*}(s,a)-Q(s,a)\big)^2 \big] \\
         &= \mathbb{E}\big[\big(r+\gamma \max_{a'} Q_{\phi,T}(s',a')-Q_{\phi}(s,a)\big)^2 \big]
\end{align}
where  $\phi$ represents the parameters of the network.

\subsection{Actor-Critic Algorithms for RL}
 The applications of RL algorithms such as Q-learning and DQN is limited only to problems with discrete action spaces. 
 Policy-based methods provide an alternative solution for continuous stochastic environments by directly optimizing the policy by taking the gradient of the objective function  with respect to the stochastic parameterized policy $\pi_{\theta}$ :
\begin{equation}
    \displaystyle \nabla_{\theta} (J(\pi_{\theta}))=\displaystyle \nabla_{\theta}\big(\mathop{\mathbb{E}}_{\tau \sim \pi_{\theta}}[R(\tau)]\big)=\mathop{\mathbb{E}}_{\tau \sim \pi_{\theta}}\left[\Big(\sum_{t=0}^{T}\nabla_{\theta}\log \pi_{\theta}(a_{t}|s_{t})\Big)R(\tau)\right]
\end{equation}
where $R(\tau)$ is the return obtained from the trajectory $\tau = \{s_0,a_0,s_1,a_1,\dots\}$. The architecture of the actor-critic algorithms is based on policy-gradient, making them amenable for continuous action spaces \cite{sutton1999policy}. Policy-based (actor) methods suffer from the drawback of high-variance estimates of the gradient and lead to slow learning. The value-based (critic) methods are an indirect method for optimizing the policy by optimizing the value function. 
Actor-critic algorithms combine the advantages of both actor-only (policy-based) and critic-only (value-based) methods and learn optimal estimates of both policy and value function.
In the actor-critic methods,  policy dictates the action based on the current state,  and the critic evaluates the action taken by the actor based on the value function estimate. The parameterized policy is then updated using the value function using the gradient ascent for improving the performance. 
 Deterministic policy gradient (DPG) proposed by Silver et al. is an actor-critic off-policy algorithm use for continuous action spaces
 \cite{silver2014deterministic}. It uses the expected gradient of the Q-value function to evaluate the gradient of the objective function with respect to parameter $\theta$ to find the optimal policy as given below:
\begin{align}
\nabla_{\theta} J(\mu_{\theta}) = \displaystyle \mathop{\mathbb{E}}[\nabla_{\theta}Q_{\mu}(s,a)|_{a=\mu_{\theta}(s)}]
\end{align}
where $\mu$ is the deterministic policy. However, they have only investigated the performance using linear function approximators to evaluate an unbiased estimate of the Q-value. 

DPG algorithm is further extended by Lillicrap et al. to  Deep deterministic policy gradient (DDPG)  \cite{Lillicrap2015Continuous} by employing DQN as a non-linear function approximator for the estimation of  Q-values. DDPG incorporate the merits of experience replay buffer and target networks to learn stable and robust Q-values. In the actor part of DDPG architecture,  rather than directly copying the weights of the policy network in the target network, the target network weights are allowed to slowly track the policy network weights to improve the stability. 
The critic part of the DDPG uses a regular DQN to find the estimate of Q-value by minimising the loss function. 

Both the value-based method and actor-critic methods suffer from the problem of overestimation bias. This problem comes due to the maximisation in the target action-value function of the loss function. Since the agent is unaware of the environment in the beginning, it needs first to estimate $Q(s,a) $ and then update them further for learning an optimal policy. Since the estimates of $Q(s,a)$ are likely to be very noisy, evaluating the maxima over value function 
does not guarantee the selection of the optimal action. If the target itself is prone to error, then the value estimates are overestimated, and this bias is then propagated further through the Bellman equation during the update.

Double Q-learning or Double DQN \cite{hasselt2010double} was an attempt to provide a solution to this problem in the value-based framework, where actions are discrete, by separating the action selection step and the estimation of the value of the action step. For example, Double DQN proposed by Hasselt et al. \cite{van2016deep} uses two estimates of the Q-value: firstly, an online network is used for the selection of action that gives the maximum Q-value, and, secondly, a target network estimates the Q-value based on this action. Fujimoto et al. \cite{fujimoto2018addressing} have proved that this overestimation-bias problem also exists in an actor-critic setting which leads to the selection of sub-optimal actions, resulting in poor policy updates. 
They have addressed this problem by introducing a variant of DQN in the actor-critic framework, the twin delayed deep deterministic policy gradient (TD3).

TD3 algorithm is an extension of the DDPG algorithm with the following modifications to address some of the lacunae of DDPG. (i) To address the overestimation bias problem, the concept of clipped double Q-learning is used wherein two Q-values are learned, and the minimum of them is used to approximate the target Q-value.  Thus, TD3 has two critic networks and corresponding critic target networks reflecting the 'twin' term in its name. 
(ii) To reduce the high variance and noisy gradients while minimising the value error per update, target networks are used to reduce error propagation by delaying the policy network update until the convergence of the Q-value. This results in less frequent policy network updates than the critic network updates, which in turn allows more stable policy updates.  
(iii) To reduce the variance in the target action values,  target policy smoothing is performed by regularisation technique where clipped noise is added to the target action obtained from the policy. 

However, none of these steps avoids the local optima that would have resulted during the training of the actor networks. Even though multi-actor ensemble essentially tries to achieve that \cite{liu2020reinforcement}, but they are essentially devoid of the advantages of the TD3 algorithm explained earlier. Hence, to achieve the best of both worlds, we propose to integrate TD3 and multi-actor methods resulting in the proposed TATD3 approach. 

\section{TATD3  - the methodology}
\label{S:3}

\begin{figure*}
     \centering
     \includegraphics[width=\linewidth]{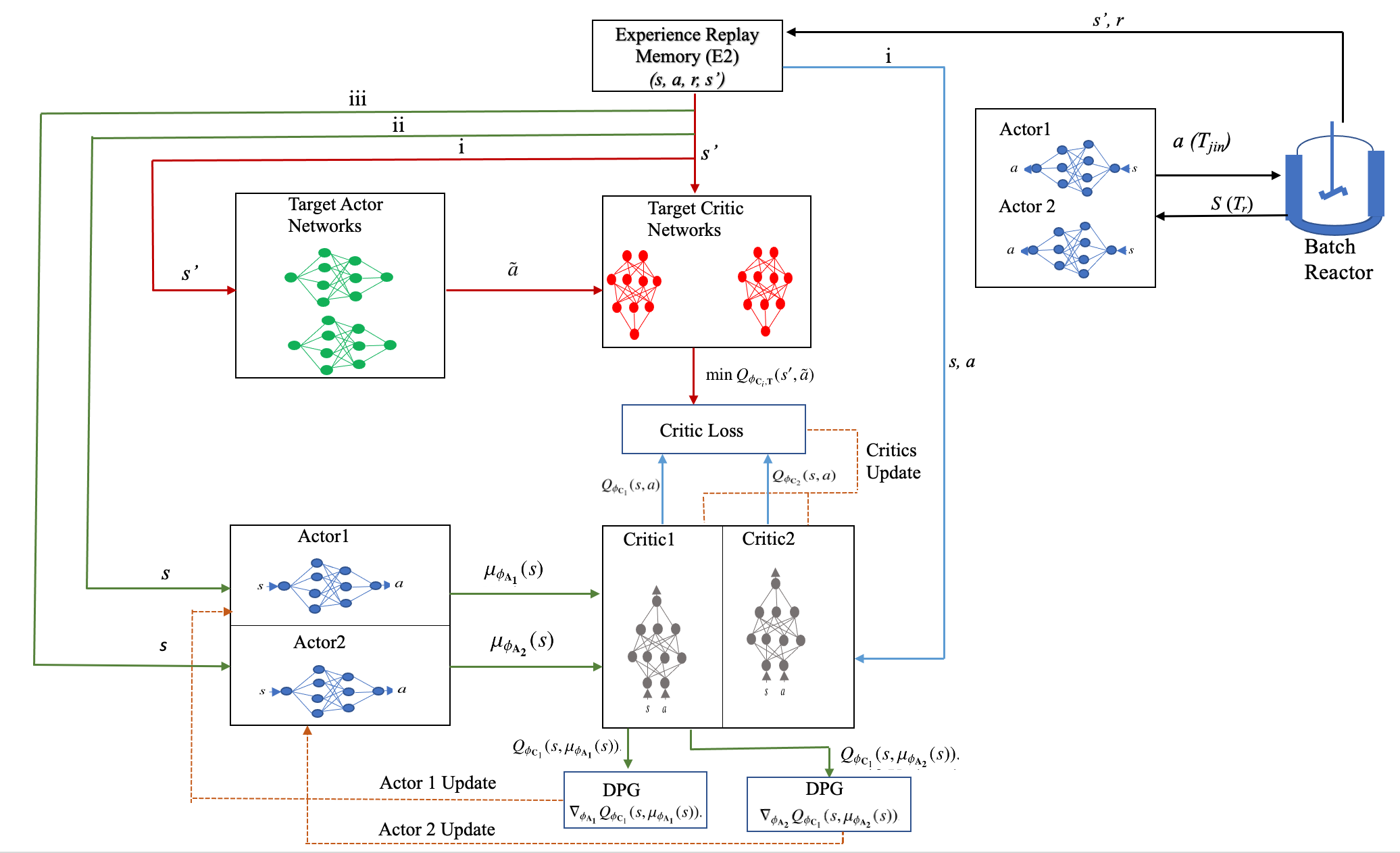}
     
    \caption{Schematic  of the TATD3 based controller for batch process}
    \label{fig:Schematic}
\end{figure*}

In this section, we present the details of the proposed TATD3 algorithm. We propose a two-stage framework for developing a TATD3 based controller for the control of the batch processes as shown in Fig. \ref{fig:Schematic}. Here in TATD3 we use twin actor-networks for policy learning instead of a single actor-network as in the vanilla TD3 algorithm. The learning of the agent (controller) includes both offline learning and online learning steps. The agent performs offline learning by interacting with the process model of the environment. 
We propose to adapt the trained actor networks (twin) and the critic networks (twin) from the offline learning step in the online learning stage, enabling a warm startup. This is expected to reduce the number of episodes required by the agent in the online stage to achieve convergence, making it suitable to apply in real-time.

Figure \ref{fig:Schematic} shows the schematic of the framework of the TATD3 based controller for the control of batch processes.
The online learning starts with the trained actor and critic models, and data obtained during the offline learning stage serves as the historical data for the agent in online learning. The historical data contains tuples of state ($s$), action ($a$), reward ($r$), next state ($s'$) which is used as the replay memory. When the TATD3 agent works in a closed-loop fashion, the two actor networks receive the initial state, $s$, from the plant and outputs the action, $a$. This action is then injected into the plant, which then reaches a new state, $s'$ and outputs the reward, $r$, for taking the selected action. We get a transition tuple, add it to the experience replay memory ($E2$).  
A batch of transition tuple, ($s$,$a$,$r$,$s'$) is randomly sampled from the experience replay memory. The state $s'$ is given to the target actor-networks which outputs the target action $\tilde{a}$. This target action and the state $s'$ are passed to the target critic networks to estimate the target Q-values. The final target value (TV) is the sum of the experience reward $r$, and the minimum discounted future reward from the critic networks.
The critic networks take the state $s$ and action, $a$, from the sampled batch of data ('$i$' in Fig. \ref{fig:Schematic}) and compute the current Q-values. The TV and the current Q- value is used to compute the loss, which is then used to update the critic networks.
To update the actor-networks a new random batch of data is sampled from the experience replay memory($E2$) for each actor ( $ii,~ iii$ in Fig. \ref{fig:Schematic}). Loss is calculated for each actor, and both the actor-networks are updated based on deterministic policy gradient (DPG).
Finally, the target network of both the critic and actor gets updated based on Polyak averaging. 
After the training, the actor-network outputs the action $a'$ for the new state $s'$, which is again injected into the plant. These steps take place in an iterative fashion until the first batch run $(b_i)$ completes. The updated model obtained for the $b_i$ batch is used for the initialisation of  networks at the start time $(t_{start})$ of the $b_{i+1}$ batch run. The whole process repeats for multiple batches runs. In this way, the agent is able to learn from the environment and thus achieve convergence. The detailed steps of the TATD3 method are mentioned in the subsection \ref{sssec:TATD3stepsinvolved}.

The following subsections contain elaborate discussions on the reward selection procedure and the detailed algorithm of TATD3.

\subsection{Reward function}
Reward function is an imperative constituent in the RL as the agent learns in the direction of increasing the reward. 
We have considered two types of reward functions inspiring the PI and PID control laws functional structure as illustrated below to accommodate the time-varying nature of the error profiles:

\begin{enumerate}
\item Reward Function 1 (PI)

\begin{align}
    r(t)=
    &\big(c_1^{\Romannum{1}}g(f(e(t)))+c_2^{\Romannum{1}}g(\sum_{t=0}^{(k-1)}f(e(t)))\big)
   \label{eq:3}
\end{align}

In Eq. \eqref{eq:3}, $c_1^{\Romannum{1}},c_2^{\Romannum{1}}$ are suitable tuning parameters,  $f(.)$ and $g(.)$ are suitable function of error $e(t):x(t)-x_{ref} \in \mathbb{R}^n$. For instance, the  function $f(.)$   may take structural forms like $||e(t)||_{w,p}$ ,   $w$ weighted $p$ norm of the vector space of $e$, while $g(.)$ is a suitable operator. Specific details regarding functions $f(.)$ and $g(.)$ are mentioned in the simulation examples.
In this reward formulation,  instead of calculating the reward only based on the current error $e(t)$, we have considered the effect of historical error profiles also.
The idea is basically to minimise the error that is being accumulated over time in the past. Since this is inspired by the principle of a  proportional-integral (PI) controller, so we are naming it as  PI Reward Function.

\item Reward Function 2 (PID)

\begin{align}
r(t)= 
&\big(c_1^{\Romannum{2}}g(f(e(t))+c_2^{\Romannum{2}}g(\sum_{t=0}^{(k-1)}f(e(t)) 
+c_3^{\Romannum{2}}g(c_4^{\Romannum{2}}+f(\Delta e(t))\big)
\label{eq:4}
\end{align}

Here, in Eq.  \eqref{eq:4}, there is an additional term  compared to Eq.  \eqref{eq:3}   $\Delta e(t):=e(t)-e(t-1)$. This particular term helps to bring in the information regarding the rate of change of error profile in the reward calculation. The {parameters} $c_1^{\Romannum{2}},\dots,c_4^{\Romannum{2}}$ are to be tuned appropriately.  Hence, we call this a proportional integral derivative (PID) reward function. In the next subsection, we proceed to illustrate simulation steps required for the implementation of the TATD3 based controller.
\end{enumerate}

\subsection{TATD3 - steps involved}
\label{sssec:TATD3stepsinvolved}
The various steps of implementation of TATD3 algorithm employed in our study are given below:

\textbf{Step 1a} Build the two actor networks and the corresponding `target actor networks' 
   with parameters $\phi_{\mathbf{A}_{1_{0}}}$, $\phi_{\mathbf{A}_{2_{0}}}$ and $\phi_{{\mathbf{A}_{1_{0}}},\mathbf{T}}$, $\phi_{{\mathbf{A}_{2_{0}}},\mathbf{T}}$, where $\mathbf{T}$, $\mathbf{A}$ denote the target network and actor network, respectively. Initialize the target actor networks as $\phi_{{\mathbf{A}_{1_{0}}},\mathbf{T}}$ $\rightarrow$  $\phi_{\mathbf{A}_{1_{0}}}$ and $\phi_{{\mathbf{A}_{2_{0}}},\mathbf{T}}$ $\rightarrow$  $\phi_{\mathbf{A}_{2_{0}}}$.
     
      \textbf{Step 1b} Build  two critic networks and initialize them with parameters $\phi_{\mathbf{C}_{1_{0}}}$ and $\phi_{\mathbf{C}_{2_{0}}}$ and the corresponding 'target critic networks' with parameters $\phi_{\mathbf{C}_{1_{0}},\mathbf{T}}$ and $\phi_{\mathbf{C}_{2_{0}},\mathbf{T}}$, where $\mathbf{C}$ denotes the critic network. Initialize the target network as $\phi_{\mathbf{C}_{1_{0}},\mathbf{T}}$ $\rightarrow$  $\phi_{\mathbf{C}_{1_{0}}}$ and $\phi_{\mathbf{C}_{2_{0}},\mathbf{T}}$ $\rightarrow$  $\phi_{\mathbf{C}_{2_{0}}}$.
      
      \textbf{Step 1c} Initialise the experience replay buffer ($E1$) set with a defined cardinality.

\textbf{Step 1d} Observe   the   initial   state, $s$   and   select   action $a$   from the actor network with noise added to the action.
     \begin{equation}
       a_i=clip(\mu_{\phi_{\mathbf{A}_{i_{0}}}}(s) + \epsilon, a_{min},a_{max}), ~i \in 
       \{1,2\}
    \end{equation}
   
    where $a_{min}$ and $a_{max}$ represent the upper and lower bounds of action, respectively, $\mu_{\phi_{\mathbf{A}_{i_{0}}}}$ is the parametrized deterministic policy.
    Further, select the  action $a$ that maximizes the $Q$ function as:
    \begin{align}
        a=\arg \underset{a_i}{\max} Q_{\phi_{\mathbf{C}_{j_{0}}}}(s,a_i)), ~i,~j \in \{1,2\}
    \end{align}
   
    \textbf{Step 1e} Execute the action, $a$ by injecting it into the process model and obtain reward $r:=r(t)$ and new state $s'$.
  
 \textbf{Step 1f} Add the obtained tuple of state, action, reward ,next state $(s,a,r:=r(t),s')$ in the replay buffer ($E1$).
    
 \textbf{Step 1g} Train both actor and critic according to \textbf{Step 6}  -\textbf{Step 13} as detailed in the online learning algorithm for $n$ desired  episodes until the convergence. 

The above steps are for offline learning. The trained models from the offline steps will be used in the online learning starting from \textbf{Step2}. Let the updated actor network parameters are denoted by $\phi_{\mathbf{A}_{1}}$ and $\phi_{\mathbf{A}_{2}}$ and the corresponding  $\mathbf{T}$ network parameters are denoted by  $\phi_{\mathbf{A},\mathbf{T_{1}}}$ and $\phi_{\mathbf{A},\mathbf{T_{2}}}$, respectively. Similarly, the updated critic model parameters are denoted by $\phi_{\mathbf{C}_{1}}$ and $\phi_{\mathbf{C}_{2}}$ and the corresponding `target critic networks' is denoted with parameters $\phi_{\mathbf{C}_{1},\mathbf{T}}$ and   $\phi_{\mathbf{C}_{2},\mathbf{T}}$ respectively.

      \textbf{Step 1} Use the trained actor and critic networks obtained from \textit{offline learning} as the actor network and the critic network for online learning.

     \textbf {Step 2}
     Initialise the experience replay buffer ($E2$) with a suitable cardinality and add tuples obtained from the offline learning  ($E1$). Each tuple is composed of a state, action, reward and new state i.e $(s,a,r,s')$.

   \textbf{Step 3} Observe the initial state $s$ and compute action $a_1$ and $a_2$ from the actor networks.  Further, it desirable to add Gaussian noise as a way of exploration during training. Also, clip the action, $a_1$ and $a_2$ between the action range, as follows:
    \begin{equation}
       a_i=clip(\mu_{\phi_{A_i}}(s) + \epsilon, a_{min},a_{max}), ~i \in 
       \{1,2\}
       \label{eq:7}
    \end{equation}
    where $\epsilon \sim \mathcal{N}(0,\,\sigma^{2})$ with a suitable exploration noise variance $\sigma^2$. Further, select the  action $a$ that maximizes the $Q$ function as:
    \begin{align}
        a=\arg \underset{a_i}{\max} Q_{\phi_{\mathbf{C}_j}}(s,a_i)), ~i,~j \in \{1,2\}
    \end{align}
   
   \textbf{Step 4} Execute the action, $a$ by injecting it into the true-process and obtain reward $r$ and new state $s'$.
    
     \textbf{Step 5} Add the obtained tuple of state, action, reward,next state $(s,a,r,s')$ in the replay buffer ($E2$).
    
      \textbf{Step 6} Sample a batch of transitions $(s,a,r,s')$ from the experience replay $E2$. 
    
   \textbf{Step 7} Each of the target actor network outputs the deterministic action  $\tilde{a_1}$, $\tilde{a_2}$  for state $s'$, subsequently, a clipped noise added to this action. The action values are further clipped to ensure that they are in the valid action range. 
    \begin{equation}
        \tilde{a_i}= clip(\mu_{\phi_{A_i,T}}(s') + clip(\epsilon,-c,c), a_{min},a_{max}), i \in\{1,2\}
    \end{equation}
    where $\epsilon \sim \mathcal{N}(0,\sigma)$; $c$ and $\sigma$ are the noise clip and the policy noise variance, respectively. Best among $\tilde{a_1}$ and $\tilde{a_2}$, $\tilde{a}$, are are chosen by the one that  maximize  $Q_\phi(s',\tilde{a})$ as below:
    
     \begin{align}
        \tilde{a}=\arg \underset{\tilde{a}_i}{\max} Q_{\phi_{\mathbf{C}_j}}(s',\tilde{a}_i)), ~i,~j \in \{1,2\}
    \end{align}
 
   \textbf{Step 8} The state $s'$ and the target action $\tilde{a} $ is given as input to the target Q-network to estimate the target Q-value, $Q_{\phi_{\mathbf{C}_1,\mathbf{T}}}(s',\tilde{a})$  and $Q_{\phi_{\mathbf{C}_2,\mathbf{T}}}(s',\tilde{a})$. Select the minimum of the two Q-values to calculate the target value ($TV$) given as:
    \begin{equation}
        TV = r+ \gamma \min Q_{\phi_{\mathbf{C}_j,\mathbf{T}}}(s',\tilde{a}), j\in \{1,2\}
    \end{equation}
    
   \textbf{Step 9}  Estimate the Q- value for the state-action pair $(s,a)$,  $Q_{\phi_{\mathbf{C}_1}}(s,a)$ and $Q_{\phi_{\mathbf{C}_2}}(s,a)$,  and calculate the Loss          (equation \eqref{EQ1}), as below,
    \begin{align}
        Loss1 &= MSE(Q_{\phi_{\mathbf{C}_1}}(s,a),TV) \\
        &= \mathbb{E}[(Q_{\phi_{\mathbf{C}_1}}(s,a)-TV)^2] \\
        Loss2 &= MSE(Q_{\phi_{\mathbf{C}_2}}(s,a),TV) \\
        &= \mathbb{E}[(Q_{\phi_{\mathbf{C}_2}}(s,a)-TV)^2] 
    \end{align}
    \begin{equation}
        Loss= MSE(Q_{\phi_{\mathbf{C}_1}}(s,a),TV) + MSE(Q_{\phi_{\mathbf{C}_2}}(s,a),TV)
    \end{equation}

   \textbf{Step 10} Update the Q-value by backpropogating the loss and update the critic network parameters $\phi_{\mathbf{C}_1}$ and $\phi_{\mathbf{C}_2}$ by stochastic gradient descent using a suitable optimizer.
    
      \textbf{Step 11} After every two iterations, randomly sample two separate sets of batch of transitions $(s,a,r,s')$ from the experience replay $E2$ for training the actor networks $A_1$ and $A_2$, respectively. This step brings in the delayed policy update.
      
   \textbf{Step 12} Update the actor networks $A_1$ and $A_2$ by doing gradient ascent on the $Q$-value of a critic  network, $\nabla_{\phi_\mathbf{A_i}} Q_{\phi_{\mathbf{C}_1}}(s,\mu_{\phi_{\mathbf{A}_i}}(s)), ~i=1,2$.

    \textbf{Step 13} Update the weights of the critic target and actor target networks. 
    
    \begin{align}
     &\text{for j= 1,2}\nonumber\\
    &  \phi_{\mathbf{C}_j,\mathbf{T}} \leftarrow \tau  \phi_{\mathbf{C}_j} + (1-\tau)\phi_{\mathbf{C}_j,\mathbf{T}} \\
   &\text{for i= 1,2}\nonumber\\
   &  \phi_{\mathbf{A_i},\mathbf{T}} \leftarrow \tau  \phi_{\mathbf{A_i}} + (1-\tau)\phi_{\mathbf{A_i},\mathbf{T}} \\
    &\text{end}\nonumber
    \end{align}
 
    by Polyak averaging, where $\tau \in [0,1] $ is the suitable target update rate.
    
     \textbf{Step 14} Obtain the new state, $s \rightarrow s'$ \& Repeat from Step 3 until the batch process completes $t=t_{end}$ where $t_{end}$ is the end time of the batch.
    
   \textbf{Step 15} Repeat  for the \textbf{Steps 3-14} for subsequent batches.

These steps are concisely presented as Algorithm 1 as next.

\begin{algorithm}
 \caption{TATD3 Algorithm}
   \begin{algorithmic}[1]
     \FOR{batch =1 , no. of batches}
       \STATE Observe the initial state s 
        \FOR{$time \in \{t_{init},\dots,t_{end}\}$}
          \STATE Compute action a (Step 3)
           \STATE Execute action $a$ to get $r$ and $s'$ 
            \STATE Add the tuple $(s,a,r,s')$ in E2 
             \FOR {$j$ in range (iteration)}
               \STATE  Randomly Sample batch of transition from E2 
                \STATE Compute the target action  as done in Step 7
                 \STATE Compute the target value (TV)
                  \STATE Update the critic network by minimising the Loss in Step9
                  \IF {j mod \emph{policy frequency}=0}
                    \FOR {Actor = 1, No.of Actors}
                      \STATE Randomly sample a  batch of transition tuples from E2\\
                      \STATE Update the Actor Network by DPG (Step 12)\\
                    \ENDFOR
                    \STATE Update target Networks of both Actor and Critic Networks(Step 13)
                 \ENDIF
              \ENDFOR
              \STATE $s$ $\rightarrow$ $s'$
             \ENDFOR
    \ENDFOR
  \end{algorithmic}
\end{algorithm}

\section{Results and Discussion}
This section discusses the numerical simulation results for the application of TATD3 based controller to two batch processes.  We have compared the performance of TATD3 with respect to other continuous RL algorithms such as TD3, DDPG, and discrete action-space RL algorithms such as DQN and Q-learning with Gaussian process (GP) \cite{rasmussen2004gaussian} as the function approximator. The algorithms, as mentioned above, are trained using two types of reward functions that resemble PI reward and PID reward, respectively. We have also introduced batch-to-batch variations and evaluated the comparative performance of  TATD3, TD3 and DDPG algorithms in this section.

\subsection{Case study 1: Batch transesterification process }

Biodiesel is produced by the transesterification reaction, wherein triglycerides (TG) from the fatty acids react with alcohols (methanol/ethanol) in the presence of a catalyst to produce methyl esters (FAME) with Diglyceride (DG) and Monoglyceride (MG) as intermediates and glycerol(GL) as a byproduct \cite{otera1993transesterification}. The lipids/fatty acids are obtained from various plant-based sources such as vegetable oil, soyabean oil, palm oil, waste cooking oil, animal fats, etc. \cite{khalizani2011transesterification,liu2008transesterification}.
 Three consecutive reversible reactions occur during this process which is given as follows: 
 
 \begin{align*}
     TG + CH_{3}OH \underset{k_2}{\stackrel{k_1}{\rightleftharpoons}} DG + R_{1}COOCH_{3} \\
     DG + CH_{3}OH \underset{k_4}{\stackrel{k_3}{\rightleftharpoons}} MG + R_{2}COOCH_{3} \\
     MG + CH_{3}OH \underset{k_6}{\stackrel{k_5}{\rightleftharpoons}} GL + R_{3}COOCH_{3} 
 \end{align*}
Here $k_{i}$ is the rate constant and is given by the Arhenius equation,  $k_{i}= ko_{i} \exp{(-E_{i}/RT_r)}$ and $T_{r}$ is the Reaction Temperature. 
The kinetic-model involving the mass balance of concentration of the species and the model assumptions is adopted from \cite{de2020constrained,noureddini1997kinetics} and is given as follows:
\begin{align*}
    \frac{d[TG]}{dt} &= -k_1[TG][A] + k_2[DG][E] \\
    \frac{d[DG]}{dt} &= k_1[TG][A] - k_2[DG][E] -k_3[DG][A] + k_4[MG][E] \\
    \frac{d[MG]}{dt} &= k_3[DG][A] - k_4[MG][E] -k_5[MG][A] + k_6[GL][E] \\
    \frac{d[E]}{dt} &= k_1[TG][A] - k_2[DG][E]+ k_3[DG][A] \\
    &
     - k_4[MG][E] +k_5[MG][A] - k_6[GL][E] \\
    \frac{d[A]}{dt} &= - \frac{d[E]}{dt} \\
    \frac{d[GL]}{dt} &= k_5[MG][A] - k_6[GL][E]\\
\end{align*}
where '[]' represents the concentration of the reactants involved and [E] is the FAME concentration.

The desired FAME concentration is mainly affected by the reactor temperature ($T_r$), and therefore this work focuses on the control of $T_r$, which can be achieved in a jacketed batch reactor by manipulating the jacket inlet temperature ($T_{jin}$).
The reactor temperature ($T_r$) and the jacket temp ($T_j$) is determined by applying an energy balance on the jacketed batch reactor. The model equations and the parameters' values are referred from  \cite{chanpirak2017improvement,kern2015advanced} and are given as follows:

\begin{align*}
    \frac{dT_r}{dt} &= \frac{M_R(-V\Delta{H}_Rr+Q_j)}{V\rho_{r}c_{m,R}} \\ 
    \frac {dT_j}{dt} &= \frac{F_j(T_{jin}-T_{j})}{V_j \rho_{j}}- \frac{Q_j}{V_j\rho_{j} c_w} \\
    Q_j &= UA(T_j -T_r)\\
    r &= \frac{d[E]}{dt}
\end{align*}

Herein, we present the training details of RL-based agents used to control the batch transesterification problem. 
A neural network consisting of 2 hidden layers with 400 and 300 hidden nodes is used for both the twin actor-networks and the twin critic-networks. Rectified Linear Unit (ReLU) is the activation function between each hidden layer for both actor networks and critic networks. Further,  a linear activation function is used for the output in the actor-networks. The network parameters are updated using the ADAM optimiser for both the actor and critics networks. Both the state and action are given as input to the critic network to estimate the Q-value. 

The implementation of the algorithm was done in Python, and the neural network framework is constructed in PyTorch (for continuous action space) and Keras (for discrete action space) API. The mathematical model of the transesterification process was simulated in Matlab and integrated into Python via the Matlab engine. Table \ref{Hyperparameters} lists the hyper-parameters used for the implementation of the TATD3 algorithm.

In this study, the  function $f(.)$ in the reward formulation (as discussed in Section 3) is taken as the absolute  value of error, $\abs{e(t)}$  i.e., $\abs{T_r(k)-T_{ref}}$ and $g(.)$ is the inverse operator ($g(.)=\frac{1}{f(.)}$). Here $T_r$ is the reactor temperature and $T_{ref}$ is the desired temperature. For the PI reward the constant values $c_1^{\Romannum{1}} , c_2^{\Romannum{1}}$ are taken as $10,100$ and for the PID reward the constant values from $c_1^{\Romannum{2}} \dots c_4^{\Romannum{2}}$ are taken as $10,100,1,0.1 $ respectively.

\begin{table}
\centering
\caption{Hyperparameters for TD3 algorithm}
\label{Hyperparameters}
\begin{tabular}{lllll}
\hline
Hyperparameters & Value 
\\
\hline
Discount factor($\gamma$)& 0.99\\
Policy noise & 0.2\\
Exploration noise & 0.1\\
Clippd noise ($c$) & 0.5\\
Actor Learning Rate & 10e-3\\
Critic Learning Rate & 10e-3\\
Target Update Rate($\tau$) &0.005 \\
Policy frequency & 2 \\
\hline
\end{tabular}
\end{table}

Table \ref{RMSE Table} compares the tracking error (in terms of RMSE values) for five different algorithms, namely, TATD3, TD3, DDPG, DQN, and GP, respectively. Neural network and Gaussian process regression (GPR) are the candidates for function approximators in DQN and GP, respectively. Further, the results are compared for the two types of reward functions considered, namely, PI and PID, as discussed in   Section \ref{S:3}. Here, the RMSE values reported are the average of the last four batches for a total of 10 batches. It can be seen that the TATD3 based controller has the lowest RMSE of 1.1502 and 1.1626 for PID and PI reward functions, respectively, as compared to TD3, DDPG and other discrete action space algorithms. 
Additionally, it can be seen that the PID reward function is a better choice for reward function due to their low RMSE for all five algorithms.

\begin{table}
\centering
\caption{Tracking error (in terms of RMSE values)  comparison  of five different  RL algorithms  for  batch transesterification process}
\label{RMSE Table}
\begin{tabular}{llllll}
\hline
\multirow{2}{*}{Reward}  & \multicolumn{2}{c}{Continuous Action} && \multicolumn{2}{c}{Discrete Action} \\
\cline{2-4}
\cline{5-6}

& \multicolumn{1}{c}{TATD3} &
\multicolumn{1}{c}{TD3} & \multicolumn{1}{c}{DDPG}& \multicolumn{1}{c}{DQN} & \multicolumn{1}{c}{GP} \\

\hline
PI & 1.1626 & 1.1785 & 1.2365 & 1.3088 & 1.3866\\
PID & 1.1502  & 1.1666 & 1.2051 & 1.2763 & 1.2875 \\
\hline
\end{tabular}
\end{table}

\begin{figure}
    \centering
    \subfloat[Continuous Action (PID)]{\includegraphics[width=0.8\linewidth]{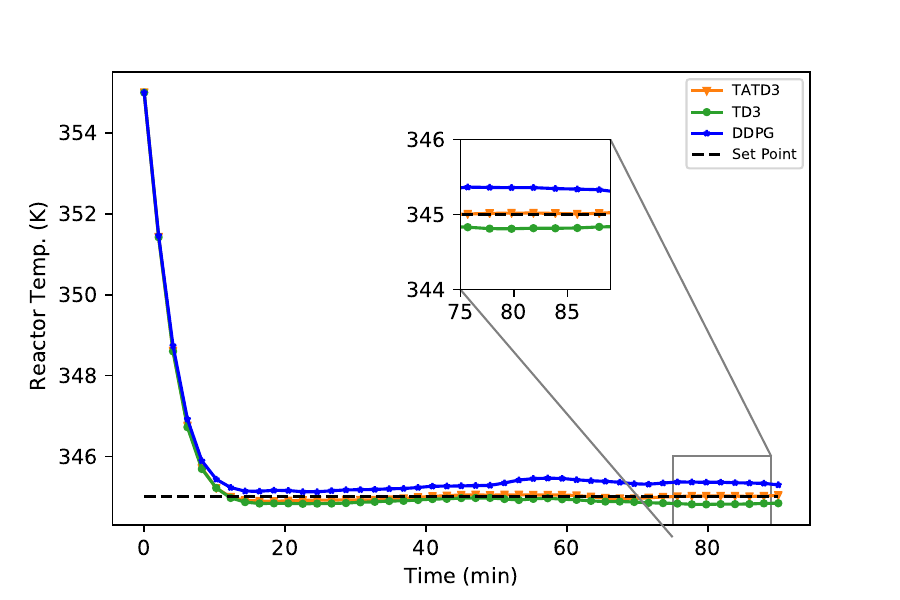}} \\
    \subfloat[Continuous Action (PI)]{\includegraphics[ width=0.8\linewidth]{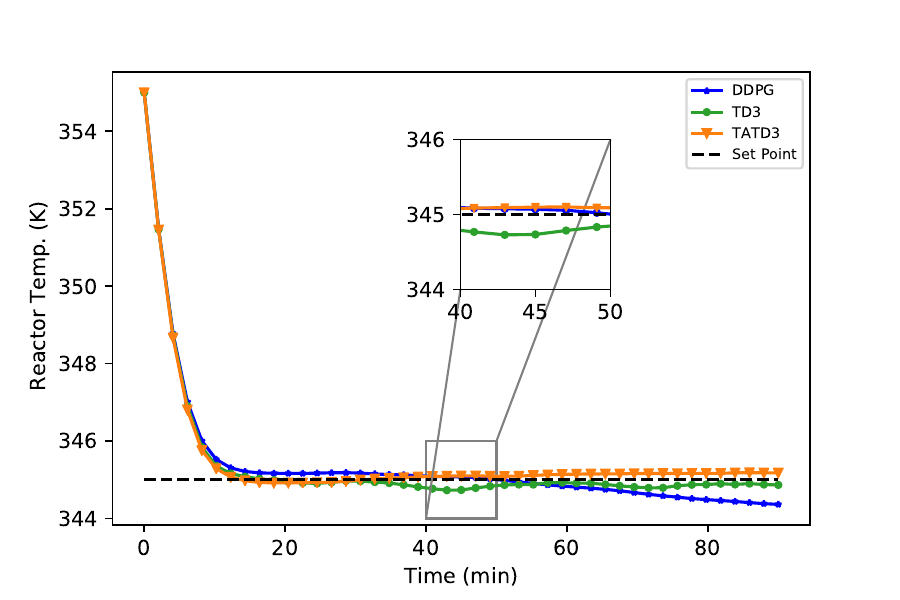}}
    \caption{Comparison of tracking performance of a) TATD3, TD3, and DDPG for PID reward and b) TATD3,TD3, and DDPG for PI reward for batch transesterification process}
    \label{fig:Response plots1_trans}
\end{figure}

\begin{figure}[ht]
    \centering
    \subfloat[Discrete Action(PID) ]{\includegraphics[width=0.8\linewidth]{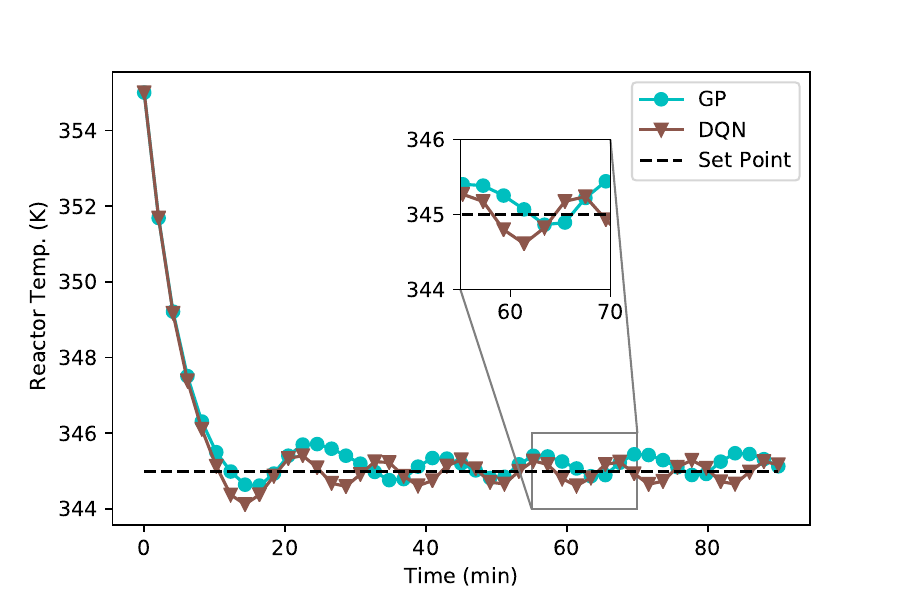}} \\
    \subfloat[Discrete Action(PI)  ]{\includegraphics[width=0.8\linewidth]{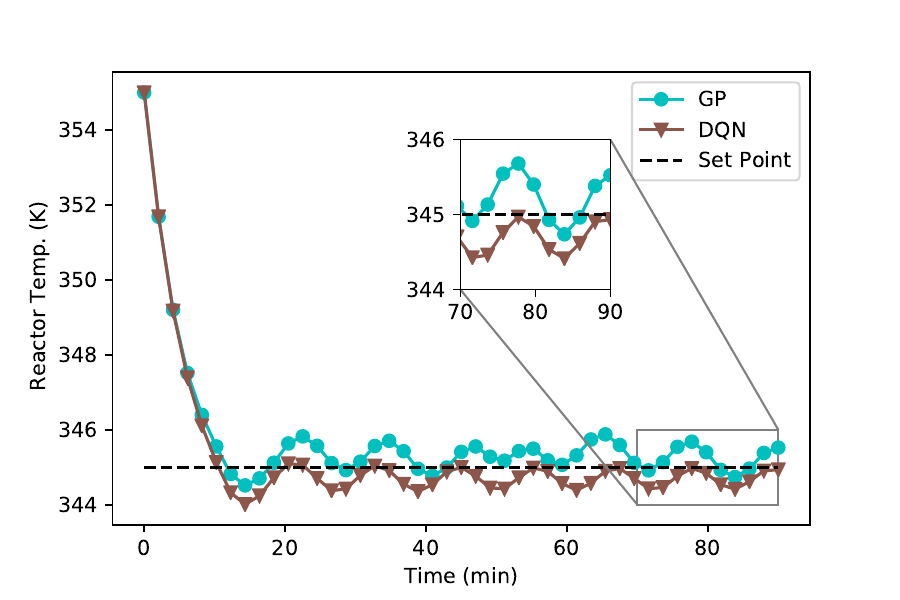}}
    \caption{ Comparison of tracking performance of a) DQN and GP for PID reward  b) DQN and GP for PI reward for batch transesterification process}
    \label{fig:Response plots2_trans}
\end{figure}

Figure \ref{fig:Response plots1_trans} and Figure \ref{fig:Response plots2_trans} shows the comparison of the tracking performance with reactor temperature ($T_r$) with respect to time. Figure \ref{fig:Response plots1_trans}  compares the tracking performance of continuous action space algorithms, namely,  TATD3 vs TD3 vs DDPG for both the PID and PI reward in subplots (a) and (b), respectively. Similarly, the performance of the discrete action space algorithms, namely, DQN and GP,  for both the reward functions are shown in Figure \ref{fig:Response plots2_trans}.
It can be seen from the plots that the reactor temperature profile closely follows the target value for TATD3 controller for the PID reward function. The results clearly conclude that that the proposed TATD3 based controller is capable of learning from the given environment and controlling the system by achieving the desired set-point of temperature $(T_{ref})$ which is 345K.

\begin{figure}[ht]
        \centering
        \subfloat[Comparison of control inputs for four different algorithms]
        {\includegraphics[width=0.8\linewidth]{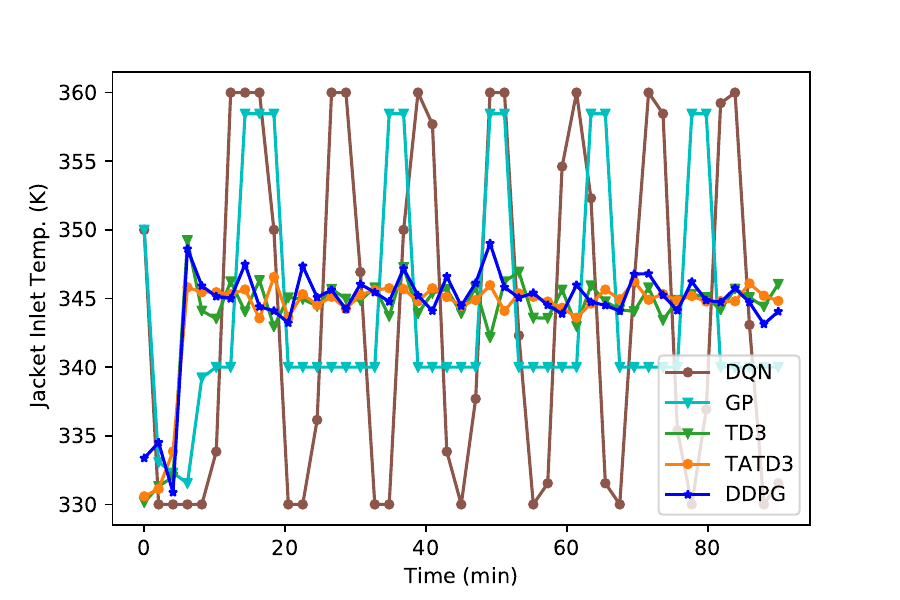}} \\
        \subfloat[Comparison of Penalty for TD3 and DDPG algorithm]
        {\includegraphics[width=0.8\linewidth]{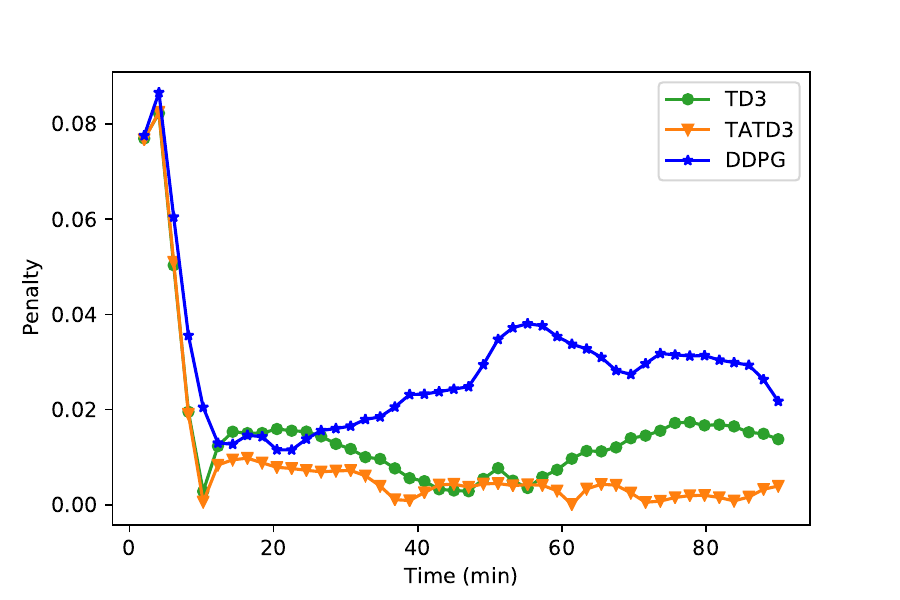}}
        \caption{Comparison of (a) control inputs  of all approaches for PID reward (b)  penalty for TATD3, TD3 and DDPG for PID reward for batch transesterification process }
        \label{fig: Graphs_trans}
\end{figure}

Since the steady-state error with the PID reward function is better than the PI reward function for all the five algorithms,  the subsequent analysis considers only the PID reward function. It can be seen that the variability in control actions is more for the discrete case as compared to the continuous action space. Table \ref{Variability in Control Actions_trans} reports the average value of standard deviation (SD) of control actions across the last four batches. The results show that the SD values for the TATD3 are less than TD3 and DDPG while the input fluctuations are more for the discrete case. 
It is worth noting that we have constrained the action space between 330-350 K for continuous action space algorithms. However, we have observed that the discrete action space algorithms are unable to honour this constraint due to limited discrete action options. Hence, we have relaxed the constraint space to 330-360 K for discrete action space algorithms, with an interval of 0.75 K.
Figure \ref{fig: Graphs_trans} (a) shows the control action plots for all the five algorithms for the PID reward.
Further, the control effort is calculated and compared for TATD3, TD3 and DDPG by taking the average of integral of the square of the control actions and the obtained values are reported in Table \ref{Computational Time_trans}. 

We have also compared the results of reward vs time plots for TATD3, TD3, and DDPG. Here, in order to make the contrast visible,  we have evaluated the inverse of the reward function,  the penalty,  and plotted the penalty values vs time as shown in Figure \ref{fig: Graphs_trans} (b).
Figure \ref{fig: Graphs_trans}(b) shows that DDPG and TD3 has a higher value of penalty than TATD3 algorithm, indicating low rewards obtained in comparison with TATD3. These results reinforce that TATD3 is better than TD3 for the control application of batch transesterification processes.

\begin{figure}[ht]
     \centering
      \includegraphics[width=0.8\linewidth]{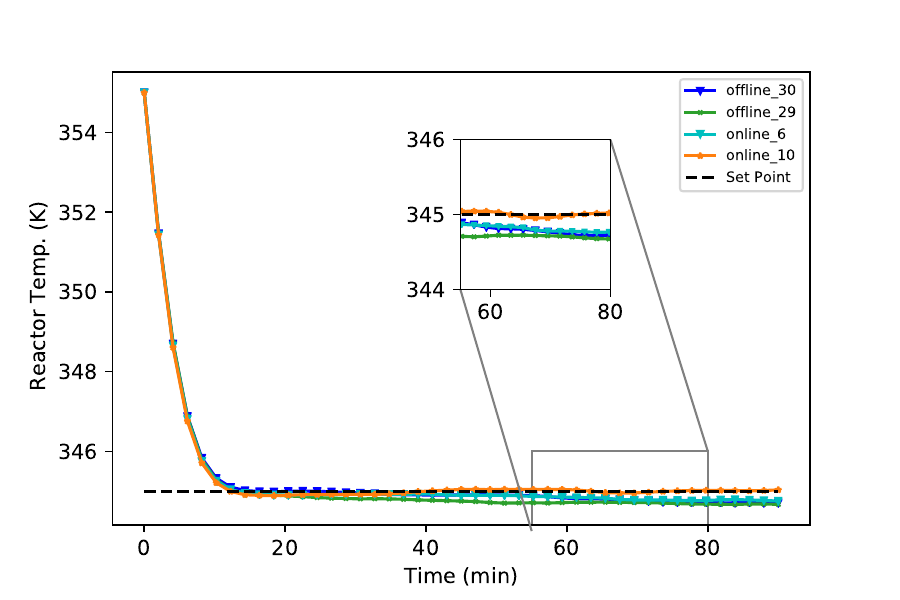}
     
     \caption{Comparison of controller performance of TATD3 offline and online learning(for batch transesterification process )}
    \label{fig:offon}
\end{figure}

To compare offline and online learning, the controller performance is compared when the agent is trained with offline learning and then followed by the online learning. The online learning starts after 30 batches of the offline learning.  It can be seen from Figure \ref{fig:offon} that the tracking performance gets gradually improved (in online learning) when the agent is trained with the network parameters learned during the offline learning part, and the best tracking performance is achieved at the  10th batch of online learning. Whereas, the offline learning has a steady state error.

 \begin{table}
\centering
\caption{Variability in control action for four different RL algorithm for the PID reward for batch transesterification process }
\label{Variability in Control Actions_trans}
\begin{tabular}{lllll}
\hline
Algorithm & SD  
\\
\hline
TATD3 &  3.3229 \\
TD3 &   3.5532\\
DDPG &  3.3726 \\
DQN &  12.068\\
GP &   10.115\\
\hline
\end{tabular}
\end{table}

\begin{table}
\centering
\caption{Comparison of control effort  for continuous action space algorithms for batch transesterification process}
\label{Computational Time_trans}
\begin{tabular}{lllll}
\hline
Algorithm & Control Effort 
\\
\hline
TATD3 & 118515.80  \\
TD3 & 118597.47\\
DDPG &  118553.28 \\
\hline
\end{tabular}
\end{table}

\subsection{Case study 2: Exothermic batch  process }

Here we consider a second-order batch exothermic chemical reaction $A\rightarrow B$. takes place in a   batch reactor having non-linear dynamics. 
The following mathematical model gives the dynamics of the batch reactor:

\begin{align*}
    \frac{dT_r}{dt} &= \frac{-UA(T_r-T_j)}{MC_p} - \frac{\Delta{H} V}{MC_p}k_o\exp{(-E/RT)}C_A^2 \\
    \frac {dC_{A}}{dt} &= -k_o\exp{(-E/RT)}C_A^2
\end{align*}
The mathematical model and the parameters are taken from \cite{lee1997model}. It is assumed that the reactor has a cooling jacket where the jacket temperature can be directly manipulated. The goal is to control the temperature of the batch reactor ($T_r$) using the coolant temperature as the manipulated variable. The action space has minimum and maximum constraint i.e.,  $290 < T_{j} < 318$ and the setpoint value $T_{ref}$ is 303K

The neural network architecture, activation function, and the actor and critic networks optimiser are the same as described in the batch transesterification process. The implementation of the algorithm was done in  Python, and the neural network framework is constructed in  PyTorch API. The mathematical model of the exothermic batch process was simulated in Matlab and integrated into Python via the Matlab engine.

Table \ref{RMSE Table2} compares the tracking error (in terms of RMSE values) for three different algorithms for continuous action space, namely, TATD3, TD3, DDPG.  Further, the results are compared for the PI and PID reward functions. Here, the RMSE values reported are the average of the last four batches for a total of 10 batches. It can be seen that the TATD3 based controller has the lowest RMSE of 0.7022 and 0.7289 for PID and PI reward functions, respectively, as compared to TD3, DDPG.
Additionally, it can be seen that the PID reward function is a better choice for reward function due to their low RMSE among all three algorithms.  
 
\begin{table}
\centering
\caption{Tracking error (in terms of RMSE values) of three different  RL algorithms  for  exothermic batch process}
\label{RMSE Table2}
\begin{tabular}{llll}
\hline
\multirow{2}{*}{Reward}  & \multicolumn{3}{c}{Continuous Action}\\
\cline{2-4}
& \multicolumn{1}{c}{TATD3} & 
\multicolumn{1}{c}{TD3}&
\multicolumn{1}{c}{DDPG}\\

\hline
PI & 0.7289  & 0.7400 & 0.7782 \\
PID & 0.7022 & 0.7324 & 0.7732 \\

\hline
\end{tabular}
\end{table}

\begin{figure}
    \centering
    \subfloat[Continuous Action (PID)]{\includegraphics[width=0.8\linewidth]{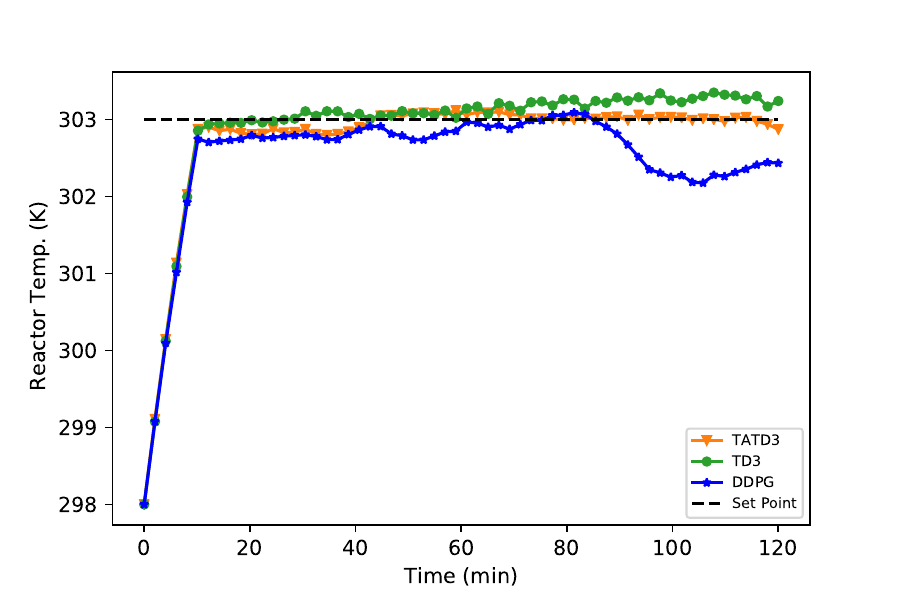}} \\
    \subfloat[Continuous Action (PI)]{\includegraphics[ width=0.8\linewidth]{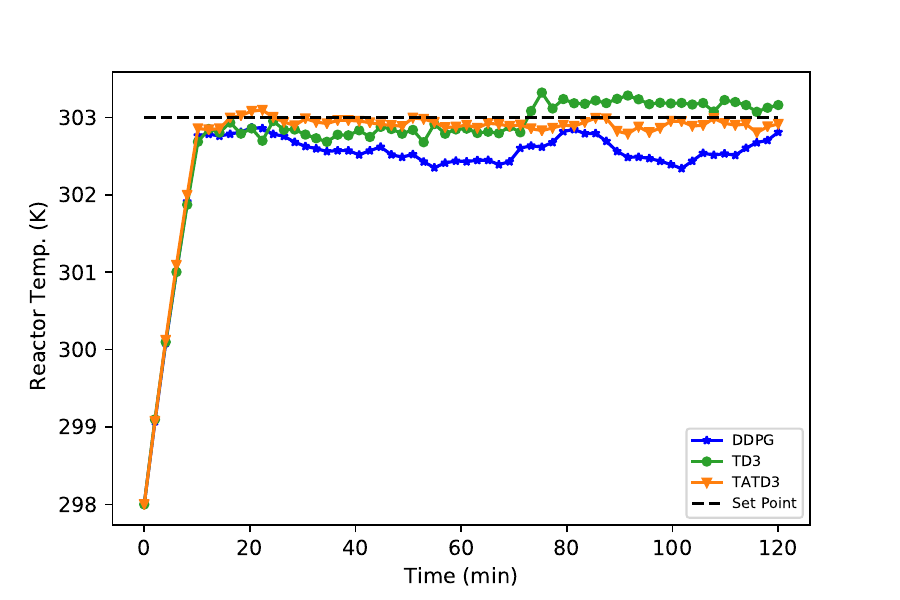}}
    \caption{Comparison of tracking performance of a) TATD3, TD3, and  DDPG for PID reward and b) TATD3,  TD3, and DDPG for PI reward for exothermic batch process}
    \label{fig:Response plots1}
\end{figure}

\begin{figure}
        \centering
        \includegraphics[width=0.8\linewidth]{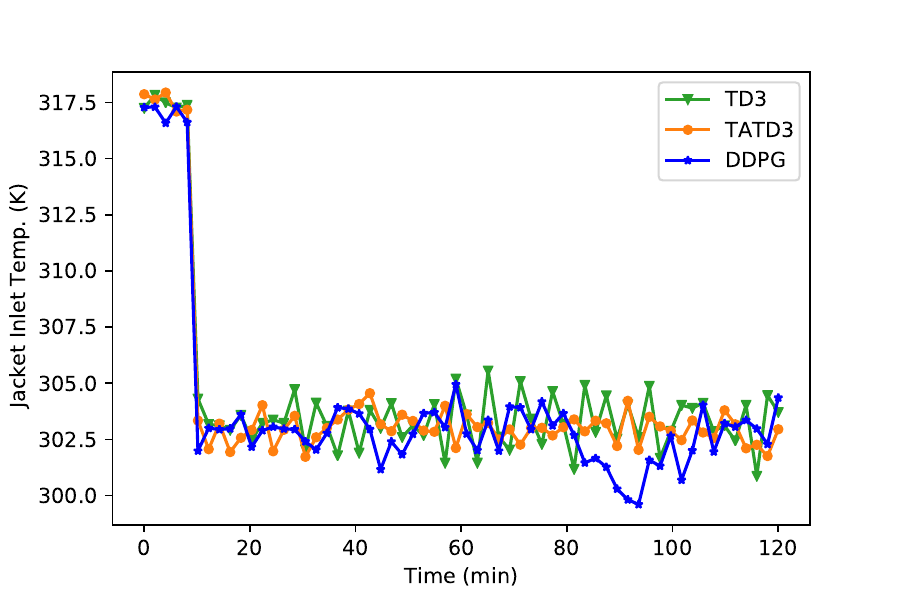}
        \caption{Control input profiles :  TATD3, TD3, and  DDPG for exothermic batch process
        }
        \label{fig:control_action_batch2}
\end{figure}

Figure \ref{fig:control_action_batch2} shows the control action plots for all the three algorithms for the PID reward showing TATD3 has less variability in control action as compared to TD3 and DDPG.
Further, the control effort is calculated and compared for TATD3, TD3 and DDPG by taking the average of integral of the square of the control actions and the obtained values are reported in Table \ref{Computational Time_batch2}. 

\begin{table}
\centering
\caption{Comparison of control effort  for continuous action space algorithms{batch exothermic process}  }
\label{Computational Time_batch2}
\begin{tabular}{lllll}
\hline
Algorithm & Control Effort 
\\
\hline
TATD3 &  92532.72 \\
TD3 & 92720.78 \\
DDPG &  92614.60 \\
\hline
\end{tabular}
\end{table}

\subsection{Effect of batch-to-batch variation}
\begin{figure}
        \centering
        \includegraphics[width=0.8\linewidth]{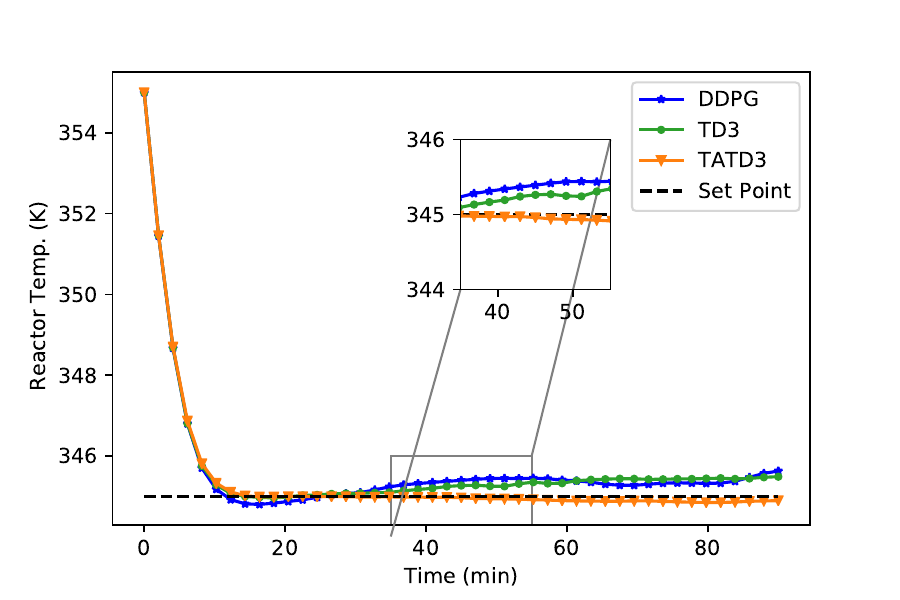}
        \caption{Comparison of tracking performance of TATD3 vs.TD3 vs. DDPG for PID reward(for batch-to-batch variation in batch transesterification process)}
        \label{fig:btbv}
\end{figure}

Further, to test the efficacy of the TATD3 controller, we have introduced batch-to-batch variations in the simulations.  Batch-to-batch variation may occur due to slight perturbations in the process parameters and changes in the environmental conditions during a  batch run. 
We have introduced batch-to-variation for the batch transesterification process by randomly changing the rate constant, $k_c$, by changing the pre-exponential factor ($k_o$) having a variance of $10\%$  in each batch. The tracking trajectory plot presented in Figure \ref{fig:btbv} clearly shows that the TATD3 based controller is able to reach the set-point in the presence of batch-to-batch variations, and thus, we achieve the desired control performance.
The average tracking error (in terms of RMSE) is 1.1564, 1.1732 and 1.2448 for TATD3, TD3, and DDPG, respectively, for batch-to-batch variations. This once again indicating the advantages of TATD3 over TD3 \& DDPG.

\section{Conclusion}
This paper proposed a control strategy of batch processes based on the TATD3 RL algorithm. The proposed algorithm is tested on two batch process case studies: (i) batch transesterification process and (ii) exothermic batch process. The reactor temperature is considered the state (control variable) to be controlled, and the jacket inlet temperature is taken as action (control input). 
It was observed that the controller is able to learn the optimal policy and achieve the desired reactor temperature by implementing appropriate control actions. We also formulated reward functions taking inspiration from the functional structure of PI and PID controller by incorporating the historical errors and showed that it helps the agent to better learn about the process.
The results indicate that TATD3 shows better convergence than continuous action-space algorithms such as TD3, DDPG, and discrete action algorithms such as DQN and GP. In summary, TATD3 based RL  controller is able to learn and intervene the process operation and control the process operation efficiently and can be used as a potential framework for complex non-linear systems where both the state and the action space are continuous.  The results indicate that the TATD3  algorithm can be a promising direction towards the goal of artificial intelligence-based control in process industries.


 \section*{Acknowledgement}
 The authors gratefully acknowledge the funding received from SERB India with the file number CRG/2018/001555.





\bibliographystyle{elsarticle-num-names}
\bibliography{sample.bib}







\end{document}